\documentclass[12pt]{article}
\usepackage{times}
\usepackage{graphicx}
\begin{document}
\begin{center}
{\Large \bf Predicting Secondary Structures, Contact Numbers, and Residue-wise Contact Orders of Native Protein Structure from Amino Acid Sequence by Critical Random Networks}  

{Akira R. Kinjo$^*$  and Ken Nishikawa

{\em Center for Information Biology and DNA Data Bank of Japan,\\
National Institute of Genetics, Mishima, 411-8540, Japan;\\
Department of Genetics, The Graduate University for Advanced Studies \\
(SOKENDAI), Mishima, 411-8540, Japan}}

\end{center}

\begin{flushleft}
Running title: Protein structure prediction in 1D.

$^*$Correspondence to A. R. Kinjo.\\
Center for Information Biology and DNA Data Bank of Japan,\\
National Institute of Genetics, Mishima, Shizuoka, 411-8540, Japan\\
Tel: +81-55-981-6859\\
Fax: +81-55-981-6889\\
E-mail: akinjo@genes.nig.ac.jp
\end{flushleft}

\begin{abstract}
Prediction of one-dimensional protein structures such as secondary structures 
 and contact numbers is useful for the three-dimensional structure 
prediction and important for the understanding of sequence-structure
 relationship. 
Here we present a new machine-learning method, critical random networks (CRNs), 
for predicting one-dimensional structures, and apply it, with position-specific
scoring matrices, to the prediction of secondary structures (SS), contact 
numbers (CN), and residue-wise contact orders (RWCO).
The present method achieves, on average, $Q_3$ accuracy of 77.8\% for SS,
correlation coefficients of 0.726 and 0.601 for CN and RWCO, respectively. 
The accuracy of the SS prediction is comparable to other state-of-the-art 
methods, and that of the CN prediction is a significant improvement over  
previous methods. We give a detailed formulation of critical random 
networks-based prediction scheme, and examine the context-dependence of 
prediction accuracies. In order to study the nonlinear and multi-body effects,
we compare the CRNs-based method with a purely linear method based on 
position-specific scoring matrices. Although not superior to the CRNs-based 
method, the surprisingly good accuracy achieved by the linear method highlights 
the difficulty in extracting structural features of higher order from amino 
acid sequence beyond that provided by the position-specific scoring matrices.
\end{abstract}
\begin{flushleft}
\end{flushleft}

\section*{Introduction}
Predicting the three-dimensional structure of a protein from its 
amino acid sequence is an essential step toward the 
thorough bottom-up understanding of complex biological phenomena.
Recently, much progress has been made in developing 
so-called \emph{ab initio} or \emph{de novo} structure prediction 
methods\cite{BonneauANDBaker2001}.
In the standard approach to such \emph{de novo} structure predictions,
a protein is represented as a physical object in three-dimensional (3D) space,
and the global minimum of free energy surface is sought with a given 
force-field or a set of scoring functions. In the minimization process, 
structural features predicted from the amino acid sequence
may be used as restraints to limit the conformational space to be sampled.
Such structural features include so-called one-dimensional (1D) structures of 
proteins.

Protein 1D structures are 3D structural features
projected onto strings of residue-wise structural assignments 
along the amino acid sequence\cite{Rost2003}. 
For example, a string of secondary structures is a 1D structure. 
Other 1D structures include (solvent) 
accessibilities\cite{LeeANDRichards1971}, contact 
numbers\cite{KinjoETAL2005} and recently introduced residue-wise contact 
orders\cite{KinjoANDNishikawa2005}. 
The contact number, also referred to as coordination number or Ooi 
number\cite{NishikawaANDOoi1980}, 
of a residue is the number of contacts that the residue makes with 
other residues in the native 3D structure, while the residue-wise 
contact order of a 
residue is the sum of sequence separations between that residue and 
contacting residues.
We have recently shown that it is possible to reconstruct the native 
3D structure of a protein from a set of three types of native 1D structures, 
namely secondary structures (SS), contact numbers (CN), and residue-wise 
contact orders (RWCO)\cite{KinjoANDNishikawa2005}. 
Therefore, these 1D structures contain rich information regarding the 
corresponding 3D structure, and their accurate prediction may be very helpful
for 3D structure prediction. 

In our previous study\cite{KinjoETAL2005}, we have developed a simple linear 
method to predict contact numbers from amino acid sequence. In that method, 
the use of multiple sequence alignment 
was shown to improve the prediction accuracy, achieving an average 
correlation coefficient of 0.63 between predicted and observed contact 
numbers per protein. There, we used amino acid frequency table obtained from
 the HSSP\cite{HSSP} multiple sequence alignment.

In this paper, we extend the previous method by introducing a new framework 
called critical random networks (CRNs), and apply it to the prediction of
secondary structure and residue-wise contact order in addition to contact 
number prediction. In this framework, a state vector of a large dimension 
is associated with each site of a target sequence.
The state vectors are connected via random nearest-neighbor interactions.
The value of the state vectors are determined by solving an equation of 
state. Then a 1D quantity of each site is predicted as a linear function
of the state vector of the site as well as the corresponding local PSSM segment.
This approach was inspired by the method of echo state networks 
(ESNs) which has been recently developed and successfully applied 
to complex time series analysis\cite{Jaeger2001,JaegerANDHaas2004}. 
Unlike ESNs which treat infinite series of input signals in one direction 
(from the past to the future), CRNs treat finite systems incorporating 
both up- and downstream information at the same time. Also, 
the so-called echo state property is not imposed to a network, 
but the system is instead set at a critical point of the network.
As the input to CRNs-based prediction, we employ position-specific 
scoring matrices (PSSMs) generated by PSI-BLAST\cite{AltschulETAL1997}. 
By the combination of PSSMs and CRNs, accurate prediction of 
SS, CN and RWCO have been achieved. 

Currently, almost all the accurate methods for one-dimensional structure 
predictions combine some kind of sophisticated machine-learning approaches 
such as neural networks and support vector machines with PSSMs. The method
presented here is no exception.
This trend raises a question as to what extent the machine-learning 
approaches are effective. In this study, we address this question by comparing
the CRNs-based method with a purely linear method based on PSSMs. Although
not so good as the CRNs-based method, the linear predictions are of 
surprisingly high quality. This result suggests that, although not 
insignificant, the effect of the machine-learning approaches is relatively 
of minor importance while the use of PSSMs is the most significant ingredient 
in one-dimensional structure prediction. The problem of how to effectively 
extract meaningful information from the amino acid sequence beyond that 
provided by PSSMs requires yet further studies.

\section*{Materials and Methods}
\subsection*{Definition of 1D structures}
\paragraph{Secondary structures (SS)}
Secondary structures were defined by the DSSP program\cite{DSSP}.
For three-state SS prediction, the simple encoding scheme was employed.
That is, $\alpha$ helices ($H$), $\beta$ strands ($E$), and other structures
(``coils'') defined by DSSP were encoded as $H$, $E$, and $C$, respectively.
For SS prediction, we introduce feature variables $(y_i^H, y_i^E, y_i^C)$ 
to represent each type of secondary structures at the $i$-th residue position,
so that $H$ is represented as $(1,-1,-1)$, $E$ as $(-1,1,-1)$, and $C$ as 
$(-1,-1,1)$.
\paragraph{Contact numbers (CN)}
Let $C_{i,j}$ represent the contact map of a protein. Usually, the contact 
map is defined so that $C_{i,j} = 1$ if the $i$-th and $j$-th residues are in 
contact by some definition, or $C_{i,j} = 0$, otherwise. As in our 
previous study, we slightly modify the definition using a sigmoid function. 
That is, 
\begin{equation}
  C_{i,j} = 1/\{1+\exp[w(r_{i,j} - d)]\}
\end{equation}
where $r_{i,j}$ is the distance between $C_{\beta}$ ($C_{\alpha}$ 
for glycines) atoms of the $i$-th and $j$-th residues, $d = 12$\AA{} is a 
cutoff distance, and $w$ is a sharpness parameter of the sigmoid function 
which is set to 3\cite{KinjoETAL2005,KinjoANDNishikawa2005}. The rather 
generous cutoff length of 12\AA{} was shown to optimize the prediction 
accuracy\cite{KinjoETAL2005}. The use of the sigmoid function enables us to 
use the contact numbers in molecular dynamics 
simulations\cite{KinjoANDNishikawa2005}.
Using the above definition of the contact map, the contact number of the
$i$-th residue of a protein is defined as
\begin{equation}
  n_i = \sum_{j:|i-j|>2}C_{i,j}. \label{eq:defcn}
\end{equation}
The feature variable $y_i$ for CN is defined as $y_i = n_i / \log L$ where 
$L$ is the sequence length of a target protein. The normalization 
factor $\log L$ is introduced because we have observed that the contact 
number averaged over a protein chain is roughly proportional to $\log L$,
and thus division by this value removes the size-dependence of predicted
contact numbers.
\paragraph{Residue-wise contact orders (RWCO)}
RWCOs were first introduced in Kinjo and Nishikawa\cite{KinjoANDNishikawa2005}.
Using the same notation as contact numbers (see above), 
the RWCO of the $i$-th residue in a protein structure is defined by 
\begin{equation}
  o_i = \sum_{j:|i-j|>2}|i-j|C_{i,j}. \label{eq:defrwco}
\end{equation}
The feature variable $y_i$ for RWCO is defined as $y_i = o_i / L$ where 
$L$ is the sequence length. Due to the similar reason as CN, the normalization
factor $L$ was introduced to remove the size-dependence of the predicted
RWCOs (the RWCO averaged over a protein chain is roughly proportional to the 
chain length).

\subsection*{Linear regression scheme}
The input to the prediction scheme we develop in this paper is a 
position-specific scoring matrix (PSSM) of the amino acid sequence of 
a target protein.
Let us denote the PSSM by $U = (\mathbf{u}_1, \cdots , \mathbf{u}_{L})$ 
where $L$ is the sequence length of the target protein and 
$\mathbf{u}_i$ is a 20-vector containing the scores of 20 types of 
amino acid residues at the $i$-th position: 
$\mathbf{u}_i = (u_{1,i}, \cdots , u_{20,i})^{t}$.

When predicting a type of 1D structures, we first predict the feature 
variable(s) for that type of 1D structures [i.e., $y_i = y_i^H$, etc. for SS,
$n_i/\log L$ for CN, and $o_i/L$ for RWCO], and then 
the feature variable is converted to the target 1D structure.
Prediction of the feature variable $y_i$ can be considered as a mapping 
from a given PSSM $U$ to $y_i$. More formally, we are going to 
establish the functional form of the mapping $F$ in $\hat{y}_{i} = F(U,i)$
where $\hat{y}_{i}$ is the predicted value of the feature variable $y_i$.
In our previous paper, we showed that CN can be predicted to a moderate 
accuracy by a simple linear regression scheme with a 
local sequence window\cite{KinjoETAL2005}. 
Accordingly, we assume that the function $F$ can be decomposed into 
linear ($F_l$) and nonlinear ($F_n$) parts: $F = F_{l} + F_{n}$. 

The linear part is expressed as 
\begin{equation}
  F_l(U,i) = \sum_{m=-M}^{M}\sum_{a=1}^{21}D_{m,a}u_{a,i+m}
\label{eq:lin}
\end{equation}
where $M$ is the half window size of the local PSSM segment around 
the $i$-th residue, and $\{D_{m,a}\}$ are the weights to be trained. 
To treat N- and C-termini separately, we introduced 
the ``terminal residue'' as the 21st kind of amino acid residue.
The value of $u_{21,i+m}$ is set to unity if $i+m<0$ or $i+m>L$, or to zero 
otherwise. The ``terminal residue'' for the central residue ($m=0$) serves 
as a bias term and is always set to unity. 

To establish the nonlinear part, we first introduce an $N$-dimensional 
``state vector'' $\mathbf{x}_i = (x_{1,i}, \cdots , x_{N,i})^{t}$ 
for the $i$-th sequence position where the dimension $N$ is a free parameter.
The value of $\mathbf{x}_i$ is determined by solving the equation of state
which is described in the next subsection. For the moment, let us assume 
that the equation of state has been solved, and denote 
the solution by $\mathbf{x}_{i}^{*}$. The state 
vector can be considered as a function of the whole PSSM $U$ 
(i.e., $\mathbf{x}_{i}^{*} = \mathbf{x}_{i}^{*}(U)$), and 
implicitly incorporates nonlinear and long-range effects. Now, the nonlinear 
part $F_n$ is expressed as a linear projection of the state vector:
\begin{equation}
  F_n(U,i) = \sum_{k=1}^{N}E_{k}x_{k,i}^{*}(U)
\label{eq:nonlin}
\end{equation}
where $\{E_{k}\}$ are the weights to be trained. 

In summary, the prediction scheme is expressed as 
\begin{equation}
    \hat{y}_{i} = \sum_{m=-M}^{M}\sum_{a=1}^{21}D_{m,a}u_{a,i+m}
+ \sum_{k=1}^{N}E_{k}x_{k,i}^{*}(U) \label{eq:pred0}
\end{equation}
Regarding $\mathbf{u}_{i-M}, \cdots, \mathbf{u}_{i+M}$ and 
$\mathbf{x}_{i}^{*}$ as independent variables, Eq. \ref{eq:pred0} reduces to
a simple linear regression problem for which the optimal weights $\{D_{m,a}\}$ 
and $\{E_k\}$ are readily determined by using a least squares method.
For CN or RWCO predictions, the predicted feature variable can be easily 
converted to the corresponding 1D quantities by multiplying by 
$\log L$ or $L$, respectively.
For SS prediction, the secondary structure $\hat{s}_i$ of the $i$-th residue
is given by $\hat{s}_i = \mathrm{arg}\max_{s\in \{H, E, C\}}y_i^s$.

\subsection*{Critical random networks and the equation of state}
We now describe the equation of state for the system of state vectors.
We denote $L$ state vectors along the amino acid sequence by
$\mathbf{X} = 
(\mathbf{x}_{1}, \cdots , \mathbf{x}_{L}) \in \mathbf{R}^{N\times L}$, 
and define a nonlinear mapping 
$g_i : \mathbf{R}^{N\times L} \to \mathbf{R}^{N}$ for $i = 1, \cdots , L$ by
\begin{equation}
  g_i(\mathbf{X}) = \tanh \left[\beta W(\mathbf{x}_{i-1}+\mathbf{x}_{i+1})+\alpha V \mathbf{u}_{i}\right]
\end{equation}
where $\beta$ and $\alpha$ are positive constants, $W$ is an $N\times N$ 
block-diagonal orthogonal random matrix, and $V$ is an $N\times 21$ random 
matrix (a unit bias term is assumed in $\mathbf{u}_i$). 
The hyperbolic tangent function ($\tanh$) is applied element-wise. 
We impose the boundary conditions as
$\mathbf{x}_0 = \mathbf{x}_{L+1} = \mathbf{0}$.
In this equation, the term containing $W$ represents nearest-neighbor 
interactions along the sequence. The amino acid sequence information is 
taken into account as an external field in the form of 
$\alpha{}V\mathbf{u}_{i}$. Next we define a mapping 
$G : \mathbf{R}^{N\times L} \to \mathbf{R}^{N\times L}$ by
\begin{equation}
  G(\mathbf{X}) = (g_{1}(\mathbf{X}), \cdots , g_{L}(\mathbf{X})).
\end{equation}
Using this mapping $G$, the equation of state is defined as 
\begin{equation}
  \mathbf{X} = G(\mathbf{X}). \label{eq:fixpoint}
\end{equation}
That is, the state vectors are determined as a fixed point of the mapping 
$G$. More explicitly, Eq. \ref{eq:fixpoint} can be expressed as 
\begin{equation}
    \mathbf{x}_{i} = \tanh \left[\beta W(\mathbf{x}_{i-1}+\mathbf{x}_{i+1})+\alpha V \mathbf{u}_{i}\right], \label{eq:eos}
\end{equation}
for $i = 1, \cdots, L$. That is, the state vector $\mathbf{x}_i$ of the site
$i$ is determined by the interaction with the state vectors of the neighboring 
sites $i-1$ and $i+1$ as well as with the `external field' $\mathbf{u}_i$ of 
the site. The information of the external field at each site is propagated
throughout the whole amino acid sequence via the nearest-neighbor interactions.
Therefore, solving Eq. (\ref{eq:eos}) means finding the state vectors that 
are consistent with the external field as well as the nearest-neighbor 
interactions, and each state vector in the obtained solution 
$\{\mathbf{x}_i\}$ self-consistently embodies the information of the whole 
amino acid sequence in a mean-field sense.

For $\beta < 0.5$, it can be shown 
that $G$ is a contraction mapping in $\mathbf{R}^{N\times L}$  
(with an appropriate norm defined therein). 
And hence, by the contraction mapping principle\cite{TakahashiNLFA}, the 
mapping $G$ has a unique fixed point independently of the strength $\alpha$ 
of the external field.
When $\beta$ is sufficiently smaller than 0.5, 
the correlation between two state vectors, say $\mathbf{x}_{i}$ 
and $\mathbf{x}_{j}$, is expected to decay exponentially as a function of 
the sequential separation $|i-j|$.
On the other hand, for $\beta > 0.5$, the number of the fixed points
varies depending on the strength of the external field $\alpha$.
In this regime, we cannot reliably solve the equation of 
state (Eq.\ref{eq:fixpoint}). 
In this sense, $\beta = 0.5$ can be considered as a critical 
point of the system $\mathbf{X}$. 
From an analogy with critical phenomena of physical 
systems\cite{Goldenfeld1992} (note the formal similarity of Eq. \ref{eq:eos} 
with the mean field equation of the Ising model), the correlation length 
between state vectors is expected to diverge, or become long when the 
external field is finite but small. 
We call the system defined by Eq. \ref{eq:eos}
with $\beta = 0.5$ a critical random network (CRN). 

The equation of state (Eq. \ref{eq:eos}) is parameterized by two random 
matrices $W$ and $V$, and consequently, so is the predicted feature variables 
$\hat{y}_{i}$. Following a standard technique of statistical 
learning such as neural networks\cite{Haykin}, we may improve the prediction 
accuracy by averaging $\hat{y}_{i}$ obtained by multiple CRNs with 
different pairs of $W$ and $V$. This averaging operation reduces the prediction
errors due to the random fluctuations in the estimated parameters.
We employ such an ensemble prediction with 10 sets of random matrices $W$ and
$V$ in the following. The use of a larger number of random matrices for 
ensemble predictions improved the prediction accuracies slightly, but the 
difference was insignificant.

\subsection*{Numerics}
Here we describe the value of the free parameters used, and a numerical 
procedure to solve the equation of state.

The half window size $M$ in the linear part of Eq. \ref{eq:pred0} is 
set to 9 for SS and CN predictions, and to 26 for RWCO prediction. 
These values are found to be optimal in preliminary studies\cite{KinjoETAL2005,
KinjoANDNishikawa2005b}.
Regarding the dimension $N$ of the state vector, we have found that $N=2000$ 
gives the best result after some experimentation, and this value is 
used throughout. Using the state vector of a large dimension as 2000, it is 
expected that various properties of amino acid sequences can be extracted and 
memorized. If the dimension is too large, overfitting may occur, but we did
not find such a case up to $N=2000$. Therefore, in principle, the state vector
dimension could be even larger (but the computational cost becomes a problem).

Each element in the $N\times 21$ random matrix $V$ in Eq. \ref{eq:eos} is 
obtained from a uniform distribution in the range [-1, 1] and the strength 
parameter $\alpha$ is set to 0.01. 
Here and in the following, all random numbers were generated by the Mersenne 
twister algorithm\cite{MersenneTwister}. 
The $N\times N$ random matrix $W$ is obtained in the following manner. 
First we generate a random block diagonal matrix $A$ whose block sizes 
are drawn from a uniform distribution of integers 2 to 20 (both inclusive), 
and the values of the block elements are drawn 
from the standard Gaussian distribution (zero mean and unit variance). 
By applying singular value decomposition, we have $A = U\Sigma V^{t}$ 
where $U$ and $V$ are orthogonal matrices and $\Sigma$ is a diagonal 
matrix of singular values.
We set $W = UV^{t}$ which is orthogonal as well as block diagonal.

To solve the equation of state (Eq. \ref{eq:eos}), we use a simple functional
iteration with a Gauss-Seidel-like updating scheme. 
Let $\nu$ denote the stage of iteration.
We set the initial value of the state vectors (with $\nu = 0$) as 
\begin{equation}
  \mathbf{x}_{i}^{(0)} = \tanh \left[\alpha V \mathbf{u}_{i}\right].\label{eq:init_eos}
\end{equation}
Then, for $i = 1, \cdots , L$ (in increasing order of $i$), we update 
the state vectors by
\begin{equation}
  \mathbf{x}_{i}^{(2\nu+1)} \gets \tanh \left[W(\mathbf{x}_{i-1}^{(2\nu+1)}+\mathbf{x}_{i+1}^{(2\nu)})+\alpha V \mathbf{u}_{i}\right].
\label{eq:feos}
\end{equation}
Next, we update them in the reverse order. That is, for $i = L, \cdots , 1$ 
(in decreasing order of $i$), 
\begin{equation}
  \mathbf{x}_{i}^{(2\nu+2)} \gets \tanh \left[W(\mathbf{x}_{i-1}^{(2\nu+1)}+\mathbf{x}_{i+1}^{(2\nu+2)})+\alpha V \mathbf{u}_{i}\right].
\label{eq:beos}
\end{equation}
We then set $\nu \gets \nu + 1$, and iterate Eqs. (\ref{eq:feos}) and (\ref{eq:beos}) until $\{\mathbf{x}_{i}\}$ converges. The convergence criterion is 
\begin{equation}
\sqrt{\sum_{i=1}^{L}\left\|\mathbf{x}_{i}^{(2\nu+2)}-\mathbf{x}_{i}^{(2\nu+1)}\right\|_{\mathbf{R}^{N}}^{2}/{NL}}<10^{-7}
\end{equation}
where $\left\|\cdot\right\|_{\mathbf{R}^{N}}$ denotes the Euclidean norm.
Convergence is typically achieved within 100 to 200 iterations for one protein.

\subsection*{Preparation of training and test sets}
We use the same set of proteins as used in our preliminary 
study\cite{KinjoANDNishikawa2005b}. In this set, there are 680 protein domains
selected from the ASTRAL database\cite{ASTRAL}, 
each of which represents a superfamily from one of all-$\alpha$, all-$\beta$,
$\alpha/\beta$, $\alpha+\beta$ or ``multi-domain'' classes of the SCOP database 
(release 1.65, December 2003)\cite{SCOP}. Conversely, each SCOP superfamily
is represented by only one of the protein domains in the data set. 
Thus, no pair of protein domains in the data set are expected to 
be homologous to each other.
For training the parameters and testing the prediction accuracy, 15-fold
cross-validation is employed. The set of 680 proteins is randomly 
divided into two groups: one consisting of 630 proteins (training set), 
and the other consisting of 50 proteins (test set). For each training set, 
the regression parameters $\{D_{m,a}\}$ and $\{E_{i}\}$ are determined, and 
using these parameters, the prediction accuracy is evaluated for the 
corresponding test set. 
This procedure was repeated for 15 times with different random divisions, 
leading to 15 pairs of training and test sets. In this way, there is some redundancy in the training and test sets although each pair of these sets share no 
proteins in common. But this raises no problem since our objective is to 
estimate the average accuracy of the predictions. A similar validation procedure was also employed by Petersen et al.\cite{PetersenETAL2000}
In total, 750 ($= 15\times 50$) proteins were tested over which 
the averages of the measures of accuracy (see below) were calculated.

\subsection*{Preparation of position-specific scoring matrix}
To obtain the position-specific scoring matrix (PSSM) of a protein, 
we conducted ten iterations of PSI-BLAST\cite{AltschulETAL1997} search
against a customized sequence database with the E-value cutoff of 
0.0005\cite{TomiiANDAkiyama2004}. The sequence database was compiled from the 
DAD database provided by DNA Data Bank of Japan\cite{DDBJ2005}, from which 
redundancy was removed by the program CD-HIT\cite{CD-HIT} with 95\% identity 
cutoff. This database was subsequently filtered by the program 
PFILT used in the PSIPRED program\cite{Jones1999}.
We use the position-specific scoring matrices (PSSM) rather than the frequency 
tables for the prediction.

\subsection*{Measures of accuracy}
For assessing the quality of SS predictions, we mainly use $Q_3$ and 
$SOV$ (the 1999 revision)\cite{SOV99}. The $Q_3$ measure quantifies the percentage of correctly predicted residues, while the $SOV$ measure evaluates the 
segment overlaps of secondary structural elements of predicted and native 
structures. Optionally, we use $Q_s$ and $Q_s^{pre}$
(with $s$ being $H$, $E$, or $C$) and Matthews' correlation coefficient $MC$. 
The $Q_s$ is defined by the percentage of correctly predicted SS type $s$ 
out of the native SS type $s$, and $Q_s^{pre}$ is defined by the percentage 
of correctly predicted SS type $s$ out of the predicted SS type $s$.

For CN and RWCO predictions, we use two measures for evaluating the prediction 
accuracy.
The first one is the correlation coefficient ($Cor$) between the observed
($n_{i}$) and predicted ($\hat{n}_{i}$) CN or RWCO\cite{KinjoETAL2005}.
The second  is the RMS error normalized by the standard deviation of the 
native CN or RWCO ($DevA$)\cite{KinjoETAL2005}.
While $Cor$ measures the quality of relative values, $DevA$ measures that of 
absolute values of the predicted CN or RWCO.

Note that the measures $Q_3$, $SOV$,  $Cor$ and $DevA$ are defined for a 
single protein chain. In practice, we average these quantities over the 
proteins in the test sets to estimate the average accuracy of prediction.
On the other hand, per-residue measures, $Q_s$, $Q_s^{pre}$ and $MC$, 
were calculated using all the residues in the test data sets, rather than 
on a per-protein basis.

\section*{Results}
\begin{figure}[htb]
\begin{center}
\includegraphics[width=7cm]{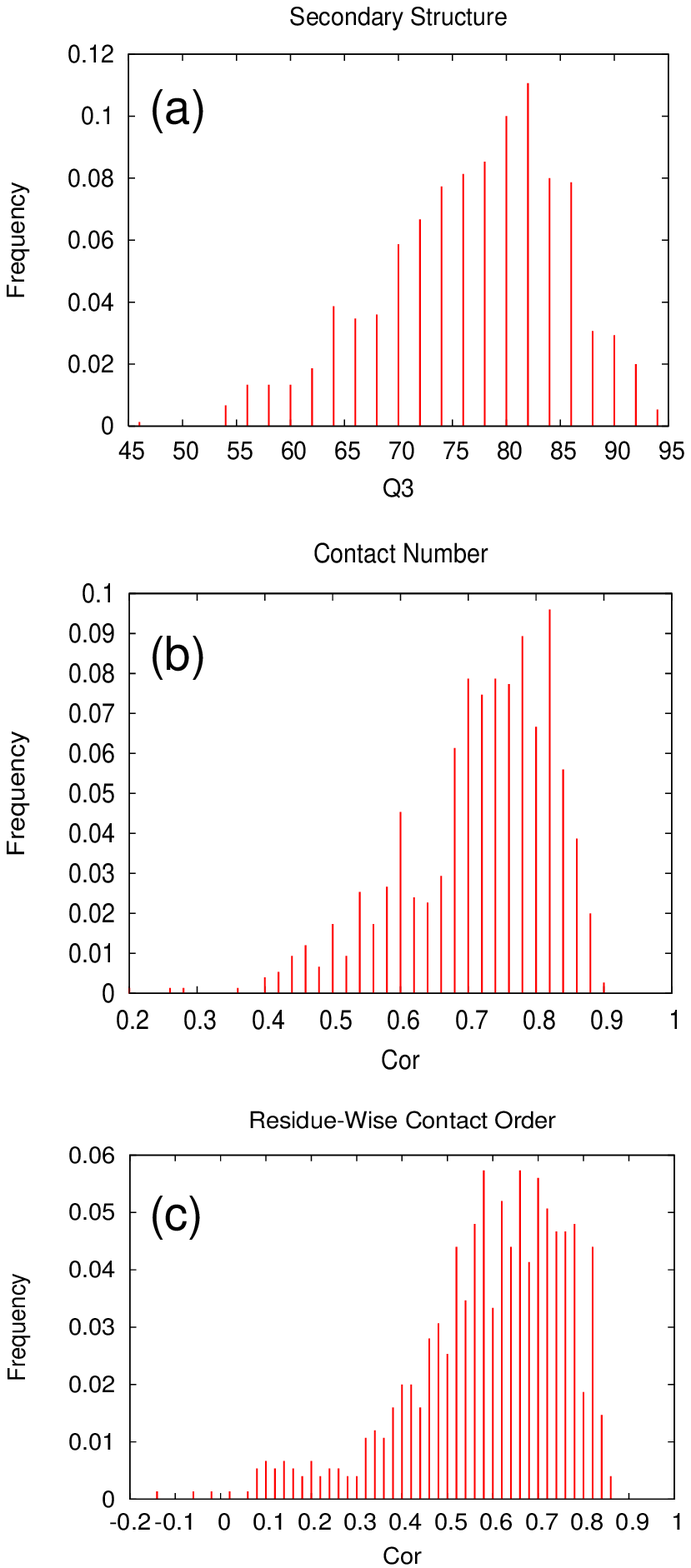}
\end{center}
\caption{\label{fig:histo}Histograms of accuracy measure obtained by  
ensemble predictions using 10 critical random networks. (a) $Q_3$ for 
secondary structure prediction; (b) $Cor$ for contact number prediction; 
(c) $Cor$ for residue-wise contact order prediction.}
\end{figure}

We examine the prediction accuracies for SS, CN, and RWCO in turn.
The main results are summarized in Table \ref{tab:summ} and Figure \ref{fig:histo}. Finally, in order to examine the effect of nonlinear terms, we verify 
the prediction results obtained using only linear terms (Eq. \ref{eq:lin}).
\begin{table}
\caption{\label{tab:summ}Summary of average prediction accuracies.}
  \begin{center}
  \begin{tabular}[h]{ll}\hline
Struct. & Accuracy \\\hline
SS  & $Q_3$ = 77.8; $SOV$ = 77.3\\
CN  & $Cor$ = 0.726; $DevA$ = 0.707\\
RWCO& $Cor$ = 0.601; $DevA$ = 0.881\\\hline
  \end{tabular}
  \end{center}
\end{table}

\subsection*{Secondary structure prediction}
The average accuracy of secondary structure prediction achieved by 
the ensemble CRNs-based approach is $Q_3=77.8$\% and $SOV=77.3$ 
(Table \ref{tab:summ}). This is comparable to the current state-of-the-art
predictors such as PSIPRED\cite{Jones1999}. The results in 
terms of per-residue accuracies ($Q_s$ and $Q_s^{pre}$) are listed in 
Table \ref{tab:ss}.
The values of $Q_s$ suggest that the present method 
underestimates $\alpha$ helices ($H$) and, especially, $\beta$ strands ($E$) 
compared to coils $C$. 
However, when a residue is predicted as being $H$ or $E$, the probability
of the correct prediction is rather high, especially for $E$ 
($Q_E^{pre} =$ 79.9\%).
The histogram of $Q_3$ (Figure \ref{fig:histo}a) shows that the peak 
of the histogram resides well beyond $Q_3$ = 80\%, and that 
only 20\% of the predictions exhibit $Q_3$ of less than 70\%. These 
observations demonstrate the capability of the CRNs-based prediction schemes.
\begin{table}
\caption{\label{tab:ss}Summary of per-residue accuracies for SS predictions.}
  \begin{center}
  \begin{tabular}[h]{lrrr}\hline
measure    & $H$ & $E$ & $C$ \\\hline
$Q_s$      & 78.4 & 61.9 & 84.6 \\
$Q_s^{pre}$ & 81.9 & 79.9 & 74.3\\
$MC$       &  0.704 & 0.636 & 0.602 \\\hline
  \end{tabular}
  \end{center}
\end{table}

\subsection*{Contact number prediction}
Using an ensemble of CRNs, a correlation coefficient ($Cor$) of 0.726 and 
normalized RMS error ($DevA$) of 0.707 was achieved for CN predictions on 
average (Table \ref{tab:summ}). This result is a significant improvement over
the previous method\cite{KinjoETAL2005} which yielded 
$Cor=0.627$ and $DevA = 0.941$. The median of the distribution of $Cor$ 
(Figure \ref{fig:histo}b) is 0.744, indicating that the majority of 
the predictions are of very high accuracy. 

We have also examined the dependence of prediction accuracy on the structural
class of target proteins (Table \ref{tab:cnhisto}). 
Among all the structural classes, $\alpha/\beta$ proteins are predicted most 
accurately with $Cor=$ 0.757 and $DevA =$ 0.668. The accuracy for other 
classes do not differ qualitatively although all-$\beta$ proteins are predicted
slightly less accurately.
\begin{table}
\caption{\label{tab:cnhisto}Summary of CN predictions for each SCOP class$^a$.}
\begin{center}
  \begin{tabular}{lrrrrr}\hline
range$^b$ &\multicolumn{5}{c}{SCOP class$^c$}\\
($Cor$) & a & b & c & d & e\\\hline
(-1,0.5]  &    8 &    6 &    3 &   14 &    1 \\
(0.5,0.6] &   19 &   25 &    8 &   19 &    1 \\
(0.6,0.7] &   29 &   29 &   22 &   54 &    3 \\
(0.7,0.8] &   62 &   66 &   76 &   85 &   10 \\
(0.8,0.9] &   43 &   38 &   57 &   67 &    3 \\
(0.9,1.0] &    1 &    0 &    0 &    1 &    0 \\
total    &  162 &  164 &  166 &  240 &   18 \\\hline
average $Cor$ & 0.721 & 0.712 & 0.757 & 0.728 & 0.722\\
average $DevA$ & 0.715 & 0.726 & 0.668 & 0.717 & 0.705\\
\hline
  \end{tabular}
\end{center}
$^a$ The number of occurrences of $Cor$ for the proteins in the test sets,
classified according to the SCOP database; average values of $Cor$ and $DevA$ 
are also listed for each class.\\
$^b$ The range ``$(x,y]$'' denotes $x < Cor \leq y$.\\
$^c$ a: all-$\alpha$; b: all-$\beta$; c: $\alpha / \beta$; d: $\alpha + \beta$;
e: multi-domain.
\end{table}

\subsection*{Residue-wise contact order prediction}
For RWCO prediction, the average accuracy was such that $Cor$ = 0.601 and 
$DevA$ = 0.881. Although these figures appear to be poor compared to those 
of the CN prediction described above, they are yet statistically significant.
The distribution of $Cor$ appears to be rather dispersed 
(Figure \ref{fig:histo}c), indicating that the prediction accuracy 
strongly depends on the characteristics of each target protein.
In a similar manner as for CN, we also examined the dependence of prediction
accuracy on the structural class of target 
proteins (Table \ref{tab:rwcohisto}).
In this case, we have found a notable dependence of prediction accuracy on 
structural classes. The best accuracy is obtained for $\alpha+\beta$
proteins with $Cor = $ 0.629 and $DevA = $ 0.832. For these proteins, 
the distribution of $Cor$ also shows good tendency in that the fraction of
poor predictions is relatively small (e.g., 14\% for $Cor <$ 0.5). 
Interestingly, all-$\beta$ proteins also show good accuracies but 
all-$\alpha$ proteins are particularly poorly predicted. These observations 
suggest that the correlation between amino acid sequence and RWCO is 
strongly dependent on the structural class of the target protein. 
However, the rather dispersed distribution of $Cor$ for each class 
(Table \ref{tab:rwcohisto}) also suggests that there are more detailed 
effects of the global context on the accuracy of RWCO prediction.
\begin{table}
\caption{\label{tab:rwcohisto}Summary of RWCO predictions for each SCOP class$^a$}
\begin{center}
  \begin{tabular}{lrrrrr}\hline
range &\multicolumn{5}{c}{SCOP class}\\
($Cor$) & a & b & c & d & e\\\hline
(-1,0.5]  &   58 &   31 &   46 &   34 &    6 \\
(0.5,0.6] &   29 &   37 &   31 &   56 &    4 \\
(0.6,0.7] &   41 &   27 &   33 &   65 &    5 \\
(0.7,0.8] &   24 &   47 &   40 &   72 &    3 \\
(0.8,0.9] &   10 &   22 &   16 &   13 &    0 \\
total    &  162 &  164 &  166 &  240 &   18 \\\hline
average $Cor$ & 0.549 & 0.620 & 0.595 & 0.629 & 0.564\\
average $DevA$ & 0.981 & 0.869 & 0.857 & 0.832 & 0.957\\
\hline
  \end{tabular}
\end{center}
$^a$See Table \ref{tab:cnhisto} for notations.
\end{table}

\subsection*{Purely linear predictions with PSSMs}
Almost all the modern methods for 1D structure prediction make use of PSSMs
in combination with some kind of machine-learning techniques such as 
feed-forward or recurrent neural networks or support vector machines. 
The present study is no exception. Curiously, machine-learning approaches 
have become so widespread that no attempt appears to have been made to test
simplest linear predictors based on PSSMs. 
In this subsection, we present results of 1D predictions using only the linear 
terms (Eq. \ref{eq:lin}) but without CRNs. In this prediction scheme, 
input is a local segment of a PSSM generated by PSI-BLAST, and a feature 
variable is predicted by a straight forward linear regression. 

As can be clearly seen in Table \ref{tab:lin},  the results of the linear 
predictions are surprisingly good although not as good as with CRNs. 
For example, in SS prediction, the purely linear scheme achieved 
$Q_3$ = 75.2\% which is lower than that of the CRNs-based scheme by only 
3.6\%. Although this is of course a large difference in a statistical sense, 
there may not be a discernible difference when individual predictions are 
concerned. (However, the improvement in the $SOV$ measure by using CRNs is 
quite large, indicating that the nonlinear terms in CRNs are indeed able to 
extract cooperative features.) It is widely accepted that the upper limit of 
accuracy ($Q_3$) of SS prediction based on a local window of a single 
sequence is less than 70\%\cite{CrooksANDBrenner2004}. 
Therefore, more than 5\% of the increase in $Q_3$ is brought simply 
by the use of PSSMs.

Similar observations also hold for CN and RWCO predictions 
(Table \ref{tab:lin}). In case of CN prediction, we have previously 
obtained $Cor$ = 0.555 by a simple linear method
 with single sequences\cite{KinjoETAL2005}. Therefore, the effect of 
PSSMs is even more dramatic than SS prediction. This may be due to the fact
that the most conspicuous feature of amino acid sequences conserved among 
distant homologs (as detected by PSI-BLAST)
is the hydrophobicity of amino acid residues\cite{KinjoANDNishikawa2004}, 
which is closely related to contact numbers.
Of course, the improvement by the use of PSSMs is largely made possible by 
the recent increase of amino acid sequence 
databases\cite{PrzybylskiANDRost2002}.
\begin{table}
\caption{\label{tab:lin}Summary of prediction accuracies using only linear terms.}
  \begin{center}
  \begin{tabular}[h]{ll}\hline
Struct. & Accuracy \\\hline
SS  & $Q_3$ = 75.2; $SOV$ = 72.7\\
CN  & $Cor$ = 0.701; $DevA$ = 0.735\\
RWCO& $Cor$ = 0.584; $DevA$ = 0.902\\\hline
  \end{tabular}
  \end{center}
\end{table}

\subsection*{The significance of criticality}
The condition of criticality ($\beta = 0.5$ in Eq. \ref{eq:eos}) is expected 
to enhance the extraction of the long-range correlations of an amino acid 
sequence, thus improving the prediction accuracy. To confirm this 
point, we tested the method by setting $\beta = 0.1$ so that the 
network of state vectors is not at the critical point any more (otherwise 
the prediction and validation schemes were the same as above). 
The prediction accuracies obtained by these non-critical random networks 
were $Q_3 = 76.7$\% and $SOV = 76.6$ for SS, 
$Cor = 0.716$ and $DevA = 0.719$ for CN, 
and $Cor = 0.589$ and $DevA = 0.897$ for RWCO.
These values are inferior to those obtained by the critical random networks
(Table \ref{tab:summ}), although slightly better than the purely linear 
predictions (Table \ref{tab:lin}). 
Therefore, compared to the non-critical random networks, the critical random 
networks can indeed extract more information from amino acid sequence and 
improve the prediction accuracies. 

\section*{Discussion}
\subsection*{Comparison with other methods}
Regarding the framework of 1D structure prediction, the critical random 
networks are most closely related to bidirectional recurrent neural
networks (BRNNs)\cite{BaldiETAL1999}, in that both can treat a whole amino 
acid sequence rather than only a local window segment. The main differences
are the following. First, network weights between input and hidden 
layers as well as those between hidden units are trained in BRNNs, 
whereas the corresponding weights in CRNs (random matrices $V$ and $W$, 
respectively, in Eq. \ref{eq:eos}) are fixed. Second, the output layer is 
nonlinear in BRNNs but linear in CRNs. Third, the network components that 
propagate sequence information from N-terminus to C-terminus are decoupled 
from those in the opposite direction in BRNNs, but they are coupled in CRNs. 

Regarding the accuracy of SS prediction, BRNNs\cite{PollastriETAL2002b} and 
CRNs exhibit comparable results of $Q_3 \approx$ 78\%. 
However, a standard local window-based approach using feed-forward neural 
networks can also achieve this level of accuracy\cite{Jones1999}. 
Thus, the CRNs-based method is not a single best predictor, but may serve as 
an addition to consensus predictions.

Although BRNNs have been also applied to CN prediction\cite{PollastriETAL2002},
 contact numbers are predicted as 2-state categorical data (buried or exposed) 
so that the results cannot be directly compared. Nevertheless, we can convert 
CRNs-based real-value predictions into 2-state predictions. By using the same 
thresholds for 
the 2-state discretization as Pollastri et al.\cite{PollastriETAL2002}
(i.e., the average CN for each residue type), 
we obtained $Q_2 =$ 75.6\% per chain (75.1\% per residue), 
and Matthews' correlation coefficient $MC =$ 0.503
whereas those obtained by BRNNs are $Q_2 =$ 73.9\% (per residue) 
and $MC =$ 0.478. 
Therefore, for 2-state CN prediction, the present method yields more 
accurate results. 

Since the present study is the very first attempt to predict RWCOs, there are 
no alternative methods to compare with. However, the comparison of 
CRNs-based methods for SS and CN predictions with other methods suggests that
 the accuracy of the RWCO prediction presented here may be the best 
possible result using any of the statistical learning methods currently 
available for 1D structure predictions. 

\subsection*{Possibilities for further improvements}
In the present study, we employed the simplest possible architecture for CRNs
in which different sites are connected via nearest-neighbor interactions.
A number of possibilities exist for the elaboration of the architecture.
For example, we may introduce short-cuts between distant sites to treat 
non-local interactions more directly. Since the prediction accuracies 
depend on the structural context of target proteins (Tables \ref{tab:cnhisto} 
and \ref{tab:rwcohisto}), it may be also useful to include  more global
features of amino acid sequences such as the bias of amino acid composition
or the average of PSSM components. These possibilities are to be pursued in 
future studies.

\section*{Conclusion}
We have developed a novel method, CRNs-based regression, for predicting
1D protein structures from amino acid sequence. When combined 
with position-specific scoring matrices produced by PSI-BLAST, this method
yields SS predictions as accurate as the best current predictors, CN
predictions far better than previous methods, and RWCO predictions 
significantly correlated with observed values. We also examined 
the effect of PSSMs on prediction accuracy, and showed that most 
improvement is brought by the use of PSSMs although the further improvement 
due to the CRNs-based method is also significant. 
In order to achieve a qualitatively yet better predictions, however, it seems
necessary to take into account other, more global, information than is 
provided by PSSMs.

\section*{Acknowledgments}
The authors thank Motonori Ota for critical comments on an early version of 
the manuscript, and Kentaro Tomii for the advice on the use of PSI-BLAST.
Most of the computations were carried out at the supercomputing facility of
National Institute of Genetics, Japan. This work was supported in part by a
grant-in-aid from the MEXT, Japan.
The source code of the programs for the CRNs-based prediction  
as well as the lists of protein domains used in this study are available at 
\verb|http://maccl01.genes.nig.ac.jp/~akinjo/crnpred_suppl/|.


\end{document}